\documentclass[12pt]{article}
\usepackage{amsfonts}

\usepackage{amsmath}

\newcommand{\bm}{\begin{multiline}}
\newcommand{\beq}{\begin{equation}}
\newcommand{\eeq}{\end{equation}}
\newcommand{\beqs}{\begin{eqnarray}}
\newcommand{\eeqs}{\end{eqnarray}}

\begin{document}

\begin{center}
\textbf{\Large NUT-Charged Black Holes in Gauss-Bonnet Gravity} \\[0pt]

\vspace{48pt}

M. H. Dehghani$^{1,2,3}$ and R. B. Mann$^{2,3}$

\vspace{12pt}

\textit{1. Physics Department and Biruni Observatory, College of
Sciences,
Shiraz University, Shiraz 71454, Iran\\[0pt]
2. Department of Physics, University of Waterloo, 200 University
Avenue
West, Waterloo, Ontario, Canada, N2L 3G1 \\[0pt]
3. Perimeter Institute for Theoretical Physics, 35 Caroline St.
N., Waterloo, Ont. Canada}
\end{center}

\begin{abstract}
We investigate the existence of Taub-NUT/bolt solutions in
Gauss-Bonnet gravity and obtain the general form of these
solutions in $d$ dimensions. We find that for all non-extremal NUT
solutions of Einstein gravity having no curvature singularity at
$r=N$, there exist NUT solutions in Gauss-Bonnet gravity that
contain these solutions in the limit that the Gauss-Bonnet
parameter $\alpha$ goes to zero. Furthermore there are no NUT
solutions in Gauss-Bonnet gravity that yield non-extremal NUT
solutions to
Einstein gravity having a curvature singularity at $r=N$ in the limit $%
\alpha \to 0$. Indeed, we have non-extreme NUT solutions in $2+2k$
dimensions with non-trivial fibration only when the
$2k$-dimensional base space is chosen to be $\mathbb{CP}^{2k}$. We
also find that the Gauss-Bonnet gravity has extremal NUT solutions
whenever the base space is a product of 2-torii with at most a
$2$-dimensional factor space of positive curvature. Indeed, when
the base space has at most one positively curved two dimensional
space as one of its factor spaces, then Gauss-Bonnet gravity
admits extreme NUT solutions, even though there a curvature
singularity exists at $r=N$. We also find that one can have bolt
solutions in Gauss-Bonnet gravity with any base space with factor
spaces of zero or positive constant curvature. The only case for
which one does not have bolt solutions is in the absence of a
cosmological term with zero curvature base space.
\end{abstract}

\vspace{30pt}

\setcounter{footnote}{0}

\thispagestyle{empty}

\hfill{}

\newpage

\section{Introduction}

Four-dimensional Taub-NUT and Taub-bolt-AdS solutions of Einstein
gravity play a central role in the construction of diverse and
interesting M-theory configurations. Indeed, the 4-dimensional
Taub-NUT-AdS solution provided the first test for the AdS/CFT
correspondence in spacetimes that are only locally asymptotically
AdS \cite{Chamblin,Hawking,MannMisner}, and the Taub-NUT metric is
central to the supergravity realization of the $D6$ brane of type
IIA string theory \cite{mbrane}. It is therefore natural to
suppose that the generalization of these solutions to the case of
Lovelock gravity, which is the low energy limit of supergravity,
might provide us with a window on some interesting new corners of
M-theory moduli space.

The original four-dimensional solution \cite{Taub,NUT} is only locally
asymptotic flat. The spacetime has as a boundary at infinity a twisted $%
S^{1} $ bundle over $S^{2}$, instead of simply being $S^{1}\times S^{2}$.
There are known extensions of the Taub-NUT solutions to the case when a
cosmological constant is present. In this case the asymptotic structure is
only locally de Sitter (for positive cosmological constant) or anti-de
Sitter (for negative cosmological constant) and the solutions are referred
to as Taub-NUT-(A)dS metrics. In general, the Killing vector that
corresponds to the coordinate that parameterizes the fibre $S^{1}$ can have
a zero-dimensional fixed point set (called a NUT solution) or a
two-dimensional fixed point set (referred to as a `bolt' solution).
Generalizations to higher dimensions follow closely the four-dimensional
case \cite{Bais,Page,Akbar,Robinson,Awad,Lorenzo,Mann1, Mann2}.

In this paper we consider Taub-NUT metrics in second order Lovelock gravity
(referred to as Gauss-Bonnet gravity), which is a higher-dimensional
generalization of Einstein gravity. In higher dimensions it is possible to
use other consistent theories of gravity with actions more general than that
of the Einstein-Hilbert action. Such an action may be written, for example,
through the use of string theory. The effect of string theory on classical
gravitational physics is usually that of modifying the low energy effective
action that describes gravity at the classical level \cite{Wit1}. This
effective action consists of the Einstein-Hilbert action plus
curvature-squared terms and higher powers as well, and in general gives rise
to fourth-order field equations containing ghosts. However if the effective
action contains the higher powers of curvature in particular combinations,
then only second-order field equations are produced and consequently no
ghosts arise \cite{Zw}. The effective action obtained by this argument is
precisely of the form proposed by Lovelock \cite{Lov}.

Until now whether or not higher derivative gravity admits
solutions of the Taub-NUT/bolt form has been an open question. Due
to the nonlinearity of the field equations, it is very difficult
to find nontrivial exact analytic solutions of Einstein's
equations modified with higher curvature terms. In most cases, one
has to adopt some approximation methods or find solutions
numerically. However, exact static spherically symmetric black
hole solutions of second and third order Lovelock gravity have
been obtained in Refs. \cite {Des,Whe,Deh1}, and of
Einstein-Maxwell-Gauss-Bonnet model in Ref. \cite{Wil}. Black hole
solutions with nontrivial topology and their thermodynamics in
this theory have been also studied \cite{Cai}. All of these
solutions in Gauss-Bonnet gravity are static. Recently two new
classes of rotating solutions of second order Lovelock gravity
have been introduced and their thermodynamics have been
investigated \cite{Deh2}.

Here we investigate the existence of Taub-NUT and Taub-bolt solutions of
Gauss-Bonnet gravity. We find that exact solutions exist, but that Lovelock
gravity introduces some features not present in higher-dimensional
Einstein-Hilbert gravity. The form of the metric function is sensitive to
the base space over which the circle is fibred. Furthermore, we find that
pure non-extreme NUT solutions only exist if the base space has a single
factor of maximal dimensionality. We conjecture that this is a general
property of Gauss-Bonnet gravity in any dimension.

The outline of our paper is as follows. We give a brief review of
the field equations of second order Lovelock gravity in Sec.
\ref{Fiel}. In Sec. \ref{6d}, we obtain all possible Taub-NUT/bolt
solutions of Gauss-Bonnet gravity in six dimensions. The structure
of these solutions suggests two conjectures, which we posit and
then check in the remainder of the paper. Then, in sections
\ref{8d} and \ref{10d}, we present all kind Taub-NUT/bolt
solutions of Gauss-Bonnet gravity in eight and ten dimensions and
check the
conjectures for them. In Sec \ref{dd}, we extend our study to the $d$%
-dimensional case, and find the general expressions for the Taub-NUT/bolt
solutions. We finish our paper with some concluding remarks.

\section{Field Equations\label{Fiel}}

The most fundamental assumption in standard general relativity is the
requirement that the field equations be generally covariant and contain at
most second order derivatives of the metric. Based on this principle, the
most general classical theory of gravitation in $d$ dimensions is Lovelock
gravity. The gravitational action of this theory can be written as
\begin{equation}
I_{G}=\int d^{d}x\sqrt{-g}\sum_{k=0}^{[d/2]}\alpha _{k}\mathcal{L}_{k}
\label{Lov1}
\end{equation}
where $[z]$ denotes the integer part of $z$, $\alpha _{k}$ is an arbitrary
constant and $\mathcal{L}_{k}$ is the Euler density of a $2k$-dimensional
manifold,
\begin{equation}
\mathcal{L}_{k}=\frac{1}{2^{k}}\delta _{\rho _{1}\sigma _{1}\cdots \rho
_{k}\sigma _{k}}^{\mu _{1}\nu _{1}\cdots \mu _{k}\nu _{k}}R_{\mu _{1}\nu
_{1}}^{\phantom{\mu_1\nu_1}{\rho_1\sigma_1}}\cdots R_{\mu _{k}\nu _{k}}^{%
\phantom{\mu_k \nu_k}{\rho_k \sigma_k}}  \label{Lov2}
\end{equation}
In Eq. (\ref{Lov2}) $\delta _{\rho _{1}\sigma _{1}\cdots \rho _{k}\sigma
_{k}}^{\mu _{1}\nu _{1}\cdots \mu _{k}\nu _{k}}$ is the generalized totally
anti-symmetric Kronecker delta and $R_{\mu \nu }^{\phantom{\mu\nu}{\rho%
\sigma}}$ is the Riemann tensor. We note that in $d$ dimensions, all terms
for which $k>[d/2]$ are identically equal to zero, and the term $k=d/2$ is a
topological term. Consequently only terms for which $k<d/2$ contribute to
the field equations. Here we restrict ourselves to the first three terms of
Lovelock gravity, which is known as Gauss-Bonnet gravity. In this case the
action is

\begin{equation}
I_{G}=\frac{1}{2}\int_{\mathcal{M}}dx^{d}\sqrt{-g}[-2\Lambda +R+\alpha
(R_{\mu \nu \gamma \delta }R^{\mu \nu \gamma \delta }-4R_{\mu \nu }R^{\mu
\nu }+R^{2})]  \label{Ig}
\end{equation}
where $\Lambda $ is the cosmological constant, $R$, $R_{\mu \nu \rho \sigma
} $, and $R_{\mu \nu }$ are the Ricci scalar and Riemann and Ricci tensors
of the spacetime, and $\alpha $ is the Gauss-Bonnet coefficient with
dimension $(\mathrm{length})^{2}$. Since it is positive in heterotic string
theory \cite{Des} we shall restrict ourselves to the case $\alpha >0$. The
first term is the cosmological term, the second term is just the Einstein
term, and the third term is the second order Lovelock (Gauss-Bonnet) term.
From a geometric point of view, the combination of these terms in
five-dimensional spacetimes is the most general Lagrangian producing second
order field equations, analogous to the situation in four-dimensional
gravity for which the Einstein-Hilbert action is the most general Lagrangian
producing second order field equations.\textbf{\ }

Varying the action with respect to the metric tensor $g_{\mu \nu }$, the
vacuum field equations are
\begin{eqnarray}
R_{\mu \nu }&-&\frac{1}{2}Rg_{\mu \nu }+\Lambda g_{\mu \nu
}-\alpha \{\frac{1}{2}g_{\mu \nu }(R_{\kappa \lambda \rho \sigma
}R^{\kappa \lambda \rho \sigma }-4R_{\rho
\sigma }R^{\rho \sigma }+R^{2})  \notag \\
&-&2RR_{\mu \nu }+4R_{\mu \lambda }R_{\text{ \ }\nu }^{\lambda
}+4R^{\rho \sigma }R_{\mu \rho \nu \sigma }-2R_{\mu }^{\ \rho
\sigma \lambda }R_{\nu \rho \sigma \lambda }\}=0  \label{Geq}
\end{eqnarray}
Equation (\ref{Geq}) does not contain derivatives of the curvature, and
therefore derivatives of the metric higher than two do not appear.

We seek Taub-NUT solutions of the field equations (\ref{Geq}). In
constructing these metrics the idea is to regard the Taub-NUT space-time as
a $U(1)$ fibration over a $2k$-dimensional base space endowed with an
Einstein-K$\ddot{a}$hler metric $g_{\mathcal{B}}$. Then the Euclidean
section of the $(2k+2)$-dimensional Taub-NUT spacetime can be written as:
\begin{equation}
ds^{2}=F(r)(d\tau +NA)^{2}+F^{-1}(r)dr^{2}+(r^{2}-N^{2})g_{\mathcal{B}}
\label{TN}
\end{equation}
where $\tau $ is the coordinate on the fibre $S^{1}$ and $A$ has a curvature
$F=dA$, which is proportional to some covariantly constant 2-form. Here $N$
is the NUT charge and $F(r)$ is a function of $r$. The solution will
describe a `NUT' if the fixed point set of the U(1) isometry $%
\partial/\partial\tau$ (i.e. the points where $F(r)=0$) is less than $2k$%
-dimensional and a `bolt' if the fixed point set is $2k$-dimensional.

\section{Six-dimensional Solutions\label{6d}}

In this section we study the six-dimensional Taub-NUT/bolt solutions (\ref
{TN}) of Gauss-Bonnet gravity. We find that the function $F(r)$ for all the
possible choices of the base space $\mathcal{B}$ can be written in the form
\begin{eqnarray}
F(r) &=&\frac{(r^{2}-N^{2})^{2}}{12\alpha (r^{2}+N^{2})}\left( 1+\frac{%
p\alpha }{(r^{2}-N^{2})}-\sqrt{B(r)+C(r)}\right)  \notag \\
B(r) &=&1+\frac{4p\alpha N^{2}(r^{4}+6r^{2}N^{2}+N^{4})+12\alpha
mr(r^{2}+N^{2})}{(r^{2}-N^{2})^{4}}  \notag \\
&&+\frac{12\alpha \Lambda (r^{2}+N^{2})}{5(r^{2}-N^{2})^{4}}%
(r^{6}-5N^{2}r^{4}+15N^{4}r^{2}+5N^{6})  \label{F6}
\end{eqnarray}
where $p$ is the sum of the dimensions of the curved factor spaces of $%
\mathcal{B}$, and the function $C(r)$\ depends on the choice of the base
space $\mathcal{B}$.

We first study the solutions for which all factor spaces of $\mathcal{B}$
are curved. The first possibility is that the base space is $\mathcal{B}=%
\mathbb{CP}^{2}$, where $A$ and the $\mathbb{CP}^{2}$ metric are:
\begin{eqnarray}
A_{2} &=&6\sin ^{2}\xi _{2}(d\psi _{2}+\sin ^{2}\xi _{1}d\psi
_{1})
\label{A2} \\
d{\Sigma _{2}}^{2} &=&6\{d{\xi _{2}}^{2}+\sin ^{2}\xi _{2}\cos ^{2}\xi
_{2}(d\psi _{2}+\sin ^{2}\xi _{1}d\psi _{1})^{2}  \notag \\
&&\ \ \ \ +sin^{2}\xi _{2}({d\xi _{1}}^{2}+\sin ^{2}\xi _{1}\cos ^{2}\xi _{1}%
{d\psi _{1}}^{2})\}  \label{CP2}
\end{eqnarray}
The function $C(r)$ for $\mathcal{B}=\mathbb{CP}^{2}$ is:

\begin{equation}
C_{\mathbb{CP}^{2}}=-\frac{16\alpha ^{2}(r^{4}+6r^{2}N^{2}+N^{4})}{%
(r^{2}-N^{2})^{4}}  \label{ac6CP}
\end{equation}

The second possibility is $\mathcal{B}=S^{2}\times S^{2}$, where $S^{2}$ is
the 2-sphere with $d\Omega ^{2}=d\theta ^{2}+\sin ^{2}\theta d\phi ^{2}$,
and the one-form $A$ is

\begin{equation}
A=2\cos \theta _{1}d\phi _{1}+2\cos \theta _{2}d\phi _{2}  \label{6SS}
\end{equation}
In this case $C(r)$ is
\begin{equation}
C_{S^{2}\times S^{2}}=-\frac{32\alpha ^{2}(r^{4}+4r^{2}N^{2}+N^{4})}{%
(r^{2}-N^{2})^{4}}  \label{c6SS}
\end{equation}
Note that the asymptotic behavior of these solutions for positive $\alpha $\
is locally flat when $\Lambda $\ vanishes, locally dS for $\Lambda >0$ and
locally AdS for $\Lambda <0$\ provided $\left| \Lambda \right| <5/(12\alpha )
$.

We note that the function $F(r)$ given in (\ref{F6}) has the same form in
the limit $\alpha \rightarrow 0$, and reduces to $F(r)$ given in \cite
{Lorenzo}. Factor spaces of zero curvature will be considered at the end of
this section.

If the base space has all factor spaces curved, then the remaining
possibilities are $\mathcal{B}=S^{2}\times H^{2}$ or $\mathcal{B}%
=H^{2}\times H^{2}$, where $H^{2}$ is the two-dimensional hyperboloid with
metric $d\Xi ^{2}=$ $d\theta ^{2}+\sinh ^{2}\theta d\phi ^{2}$. However we
find that the function $F(r)$ is not real for all values of $r$ in the range
$0\leq r\leq \infty $ for positive values of $\alpha $. The only case for
which a black hole solution can be obtained is when $\alpha $ and $\Lambda $
are negative. As we mentioned earlier, we are interested in positive values
of $\alpha $, and so we don't consider them here.

\subsection{Taub-NUT Solutions}

The solutions given in Eq. (\ref{F6}) describe NUT solutions, if (i) $%
F(r=N)=0$ and (ii) $F^{\prime }(r=N)=1/(3N)$. The first condition comes from
the fact that all the extra dimensions should collapse to zero at the fixed
point set of $\partial/\partial\tau $, and the second one ensures that there
is no conical singularity with a smoothly closed fiber at $r=N$. Using these
conditions, one finds that Gauss-Bonnet gravity in six dimensions admits NUT
solutions with a $\mathbb{CP}^{2}$ base space when the mass parameter is
fixed to be
\begin{equation}
m_{n}=-\frac{16}{15}N(3\Lambda N^{4}+5N^{2}-5\alpha )  \label{mn6cp}
\end{equation}

Computation of the Kretschmann scalar at $r=N$ for the solutions given in
the last section shows that the spacetime with $\mathcal{B}=S^{2}\times
S^{2} $ has a curvature singularity at $r=N$ in Einstein gravity, while the
spacetime with $\mathcal{B}=\mathbb{CP}^{2}$ has no curvature singularity at
$r=N$. Thus, we conjecture that ``\textit{For all non-extremal NUT solutions
of Einstein gravity having no curvature singularity at }$r=N$\textit{, there
exist NUT solutions in Gauss-Bonnet gravity that contain these solutions in
the limit that the parameter }$\alpha $\textit{\ vanishes. Furthermore there
are no NUT solutions in Gauss-Bonnet gravity that yield non-extremal NUT
solutions to Einstein gravity having a curvature singularity at }$r=N$%
\textit{\ in the limit }$\alpha \rightarrow 0$\textit{\ .}'' Indeed, we have
non-extreme NUT solutions in $2+2k$ dimensions with non-trivial fibration
when the $2k$-dimensional base space is chosen to be $\mathbb{CP}^{2}$.

We will test this conjecture throughout the rest of the paper, and consider
the case of extremal NUT solutions at the end of this section.

\subsection{Taub-Bolt Solutions}

The conditions for having a regular bolt solution are (i) $F(r=r_{b})=0$ and
(ii)$\ F^{\prime }(r_{b})=1/(3N)$ with $r_{b}>N$. Condition (ii), which
again follows from the fact that we want to avoid a conical singularity at
the bolt, together with the fact that the period of $\tau $ will still be $%
12\pi N$ , gives the following equation for $r_{b}>N$ with the base space $%
\mathbb{CP}^{2}$
\begin{equation}
3N\Lambda {r_{b}}^{3}+(2+3\Lambda N^{2}){r_{b}}^{2}-N(4+3\Lambda
N^{2})r_{b}-3\Lambda N^{4}-6N^{2}+8\alpha =0  \label{bmat6cp}
\end{equation}
which has at least one real solution. This real solution for $N<\sqrt{\alpha
}$ yields $r_{b}>N$, which is a bolt solution, while for $N\geq \sqrt{\alpha
}$, there is no bolt solution -- only the NUT solution can satisfy the
regularity conditions.

As $\Lambda $ goes to zero this real solution for $r_{b}$ diverges. However
setting $\Lambda =0$ we obtain the asymptotically locally flat case, and Eq.
(\ref{bmat6cp}) becomes a quadratic equation with the following two roots:
\begin{equation}
r_{b}=N\pm 2\sqrt{(N^{2}-\alpha )}  \label{rb6}
\end{equation}
Note that we have only one possible bolt solution (provided $N>\sqrt{\alpha}$%
), since the smaller root is less than $N$.

For the case of $\mathcal{B}=S^{2}\times S^{2}$, $r_{b}$ can be found by
solving the following equation:
\begin{equation}
3N\Lambda {r_{b}}^{4}+2{r_{b}}^{3}-6N(\Lambda N^{2}+1){r_{b}}%
^{2}-2(N^{2}-4\alpha )r_{b}+3\Lambda N^{5}+6N^{3}-12\alpha N=0
\label{bmat6s}
\end{equation}
where the condition for having bolt solution(s) is that $N\leq N_{\mathrm{%
\max }}$, where $N_{\mathrm{\max }}$ is the smaller root of the following
equation:
\begin{eqnarray}
&&-144\Lambda ^{2}(1+3\Lambda \alpha )(4+15\Lambda \alpha
)N^{10}-144\Lambda (72\Lambda ^{2}\alpha ^{2}+49\Lambda \alpha
+8)N^{8}
\notag \\
&&+8(2916\Lambda ^{3}\alpha ^{3}+2331\Lambda ^{2}\alpha ^{2}+396\Lambda
\alpha -8)N^{6}-48\alpha (216\Lambda ^{2}\alpha ^{2}+51\Lambda \alpha
-4)N^{4}  \notag \\
&&+3\alpha ^{2}(1296\Lambda ^{2}\alpha ^{2}+360\Lambda \alpha
-55)N^{2}+64\alpha ^{3}=0  \label{R6s}
\end{eqnarray}
Note that the above equation (\ref{R6s}) reduces to the equation for $N_{\mathrm{%
\max }}$ of Einstein gravity when $\alpha $ vanishes. For $N<N_{\mathrm{\max
}}$ with negative $\Lambda $, we have two bolt solutions, while for $N=N_{%
\mathrm{\max }}$\ we have only one bolt solution. It is straightforward to
show that the value of $N_{\mathrm{\max }}$ is larger in Gauss-Bonnet
gravity with respect to its value in Einstein gravity. As $\Lambda
\rightarrow 0$, the larger bolt solution goes to infinity, so for the
asymptotically flat solution, there is always one and only one bolt solution
that is the real root of the following equation:
\begin{equation}
{r_{b}}^{3}-3N{r_{b}}^{2}-(N^{2}-4\alpha )r_{b}+3N^{3}-6\alpha N=0
\label{6dbolts2s2}
\end{equation}
Eq. (\ref{6dbolts2s2}) has either one or three real roots larger
than $N$ since its left hand side is negative at $r_{b}=N$, and
positive as $r_{b}$ goes to infinity. However it cannot have three
real roots larger than $N$. This is because the product of the
roots is equal to $-3N(N^{2}-2\alpha )< 0$, whereas the condition
for having three real roots guarantees that
$N^{2}>2\alpha $. Thus there exists one and only one real root larger than $N$%
, and so there is always one bolt solution. This kind of argument applies
for all of the solutions we obtain in any dimension with any base space.

Finally, we note an additional regularity condition not present in
Einstein gravity. The metric function $F(r)$ is real for all
values of $r$ in the range $0\leq r\leq \infty $ provided $\alpha
<\alpha _{\mathrm{\max }}$, where $\alpha _{\mathrm{\max }}$
depends on the parameters of the metric. This is a general feature
of solutions in Gauss-Bonnet gravity and also occurs for static
solutions \cite{Des}.

We also note that for $\Lambda>0$ the bolt radius increases with increasing $%
\alpha$, while for the case of $\Lambda=0$ it decreases as $\alpha$
increases. For the case of $\Lambda<0$, there are two bolt solutions
provided $N<N_{\mathrm{\max }}$. As $\alpha$ increases the radius of smaller
one decreases, while that of the larger solution increases. For $N=N_{%
\mathrm{\max }}$, only the smaller bolt solution exists, and its radius $r_b$
decreases as $\alpha$ increases. These features happen for all the bolt
solutions in any dimension.

\subsection{Taub-NUT/Bolt Solutions with $T^{2}$ in the base}

We now consider the Taub-NUT/bolt solutions of Gauss-Bonnet gravity when $%
\mathcal{B}$ contains a 2-torus $T^{2}$ with metric $d\Gamma
=d\eta ^{2}+d\zeta ^{2}$. There are two possibilities. The first
possibility is to choose the base space $\mathcal{B}$ to be
$S^{2}\times T^{2}$ with 1-form
\begin{equation}
A=2\cos \theta d\phi +2\eta d\zeta  \label{TS6}
\end{equation}
and the second is $T^{2}\times T^{2}$ with 1-form
\begin{equation}
A=2\eta _{1}d\zeta _{1}+2\eta _{2}d\zeta _{2}  \label{TT6}
\end{equation}
The function $F(r)$ for the above two cases is given by Eq. (\ref{F6}),
where $C(r)$ is zero for $\mathcal{B}=T^{2}\times T^{2}$ and is
\begin{equation}
\text{\ }C_{ST}=\frac{4\alpha ^{2}}{(r^{2}-N^{2})^{2}}  \label{cst6}
\end{equation}
for $\mathcal{B}=S^{2}\times T^{2}$. The topology
$\mathcal{B}=T^{2}\times T^{2}$ at $r\rightarrow \infty $ is
$\mathbb{R}^{5}$. Although the boundary is topologically a direct
product of the Euclidean time line and the spatial hypersurface
($\eta _{1},\eta _{2},\zeta _{1},\zeta _{2}$), the product is
twisted and the boundary is not flat. Unlike the $B=CP^{2}$ case,
an immediate consequence of the $T^{2}\times T^{2}$\ topology is
that the Euclidean time period\textbf{\ }$\beta =4\pi /F^{\prime
}$\textbf{\ }will not be fixed by the value of the nut
parameter\textbf{\ }$N$\textbf{. }

Now we consider the NUT solutions. For these two cases the conditions of
existence of NUT solutions are satisfied provided the mass parameter is
\begin{eqnarray}
m_{N} &=&-\frac{16}{5}\Lambda N^{5},  \label{mtt6} \\
m_{N} &=&-\frac{8}{15}N^{3}(6\Lambda N^{2}+5)  \label{mst6}
\end{eqnarray}
for $\mathcal{B}=T^{2}\times T^{2}$ and $\mathcal{B}=T^{2}\times S^{2}$
respectively. Indeed for these two cases $F^{\prime }(r=N)=0$, and therefore
the NUT solutions should be regarded as extremal solutions. Computing the
Kretschmann scalar, we find that there is a curvature singularity at $r=N$
for the spacetime with $\mathcal{B}=T^{2}\times S^{2}$, while the spacetime
with $\mathcal{B}=T^{2}\times T^{2}$ has no curvature singularity at $r=N$.

This leads us to our second conjecture: ``\textit{Gauss-Bonnet gravity has
extremal NUT solutions whenever the base space is a product of 2-torii with
at most one }$2$\textit{-dimensional space of positive curvature}''. Indeed,
when the base space has at most one two dimensional curved space as one of
its factor spaces, then Gauss-Bonnet gravity admits an extreme NUT solution
even though there exists a curvature singularity at $r=N$.

Next we consider the Taub-bolt solutions. Euclidean regularity at the bolt
requires the period of $\tau $ to be
\begin{equation}
\beta =\frac{8\pi r_{b}(r_{b}^{2}-N^{2}+2\alpha )}{(r_{b}^{2}-N^{2})[1-%
\Lambda (r_{b}^{2}-N^{2})]}  \label{betst6}
\end{equation}
for $\mathcal{B}=T^{2}\times S^{2}$, and
\begin{equation}
\beta =-\frac{8\pi r_{b}}{\Lambda (r{_{b}}^{2}-N^{2})}  \label{bettt6}
\end{equation}
for $\mathcal{B}=T^{2}\times T^{2}$. As $r_{b}$ varies from $N$ to infinity,
one covers the whole temperature range from $0$ to $\infty $, and therefore
we have non-extreme bolt solutions. Note that for the case of $\Lambda =0$,
there is no black hole solution with $\mathcal{B}=T^{2}\times T^{2}$. This
is also true for spherically symmetric solutions of Gauss-Bonnet gravity
\cite{Cai}.

\section{Eight-dimensional Solutions\label{8d}}

In eight dimensions there are more possibilities for the base space $%
\mathcal{B}$. It can be a 6-dimensional space, a product of three
2-dimensional spaces, or the product of a 4-dimensional space with a
2-dimensional one. In all of these cases the form of the function $F(r)$ is
\begin{eqnarray}
F(r) &=&\frac{(r^{2}-N^{2})^{2}}{8\alpha (5r^{2}+3N^{2})}\left( 1+\frac{%
4p\alpha }{3(r^{2}-N^{2})}-\sqrt{B(r)+C(r)}\right)   \notag \\
B(r) &=&1-\frac{16\alpha mr\left( 5r^{2}+3N^{2}\right) }{3(r^{2}-N^{2})^{5}}+%
\frac{16p\alpha N^{2}}{15(r^{2}-N^{2})^{5}}%
(r^{6}-15N^{2}r^{4}-45N^{4}r^{2}-5N^{6})  \notag \\
&&+\frac{16\alpha \Lambda \left( 5r^{2}+3N^{2}\right) }{105(r^{2}-N^{2})^{5}}%
(5r^{8}-28N^{2}r^{6}+70N^{4}r^{4}-140N^{6}r^{2}-35N^{8})  \label{F8}
\end{eqnarray}
where $p$\ is again the dimension of the curved factor spaces of $\mathcal{B}
$, and the function $C(r)$\ depends on the choice of the base space.

The 1-form and the metric for the factor spaces $S^{2}$, $T^{2}$, and $%
\mathbb{CP}^{2}$ have been introduced in the last section. Here, for
completeness we write down the metric and 1-form for the factor space $%
\mathbb{CP}^{3}$, and then we bring the function $F(r)$ for the various base
spaces in a table. The metric and the 1-form $A$ for $\mathbb{CP}^{3}$ may
be written as:
\begin{eqnarray}
d{\Sigma _{3}}^{2} &=&8\{d{\xi _{3}}^{2}+\sin ^{2}\xi _{3}\cos ^{2}\xi
_{3}(d\psi _{3}+\frac{1}{6}A_{2})^{2}+\frac{1}{6}\sin ^{2}\xi _{3}d{%
\Sigma _{2}}^{2}\}  \label{CP3} \\
A_{3} &=&8\sin ^{2}\xi _{3}\left\{ d\psi _{3}+\sin ^{2}\xi
_{2}(d\psi _{2}+\sin ^{2}\xi _{1}d\psi _{1})\right\}  \label{A3}
\end{eqnarray}
The function $C(r)$ for various base spaces are: \bigskip

\begin{tabular}{|c|c|}
\hline
$\mathcal{B}$ & $(r^{2}-N^{2})^{5}C(r)/\alpha ^{2}$ \\ \hline
${\,{\mathbb{CP}^{3}}}$ & $-16(r^{6}-15N^{2}r^{4}-45N^{4}r^{2}-5N^{6})$ \\
\hline
$T^{2}\times T^{2}\times T^{2}$ & $0$ \\ \hline
$T^{2}\times T^{2}\times S^{2}$ & $\frac{64}{9}(r^{2}-N^{2})^{3}$ \\ \hline
$T^{2}\times S^{2}\times S^{2}$ & $-\frac{64}{9}%
(r^{6}-15N^{2}r^{4}-45N^{4}r^{2}-5N^{6})$ \\ \hline
$S^{2}\times S^{2}\times S^{2}$ & $-\frac{128}{3}%
(r^{6}-9N^{2}r^{4}-21N^{4}r^{2}-3N^{6})$ \\ \hline
$T^{2}\times \mathbb{CP}^{2}$ & $\frac{128}{27}%
(r^{6}+9N^{2}r^{4}+51N^{4}r^{2}+3N^{6})$ \\ \hline
$S^{2}\times \mathbb{CP}^{2}$ & $-\frac{64}{27}%
(13r^{6}-135N^{2}r^{4}-345N^{4}r^{2}-45N^{6})$ \\ \hline
\end{tabular}
\newline
\bigskip

One may note that the asymptotic behavior of all of these solutions is
locally AdS for $\Lambda <0$ provided $\left| \Lambda \right| <21/(80\alpha
) $, locally dS for $\Lambda >0$ and locally flat for $\Lambda =0$.

Note that all the different $F(r)$'s given in this section have the same
form as $\alpha $ goes to zero, reducing to the solutions of Einstein
gravity.

\subsection{Taub-NUT Solutions}

Using the conditions for NUT solutions, (i) $F(r=N)=0$ and (ii) $F^{\prime
}(r=N)=1/(4N)$, we find that Gauss-Bonnet gravity in eight dimensions admits
non-extreme NUT solutions only when the base space is chosen to be $\mathbb{%
CP}^{3}$. The conditions for a nonsingular NUT solution are satisfied
provided the mass parameter is fixed to be
\begin{equation}
m_{N}=-\frac{8N^{3}}{105}(16\Lambda N^{4}+42N^{2}-105\alpha )  \label{mcp3}
\end{equation}
On the other hand, the solutions with $\mathcal{B}=T^{2}\times T^{2}\times
T^{2}=\mathcal{B}_{A}$ and $\mathcal{B}=T^{2}\times T^{2}\times S^{2}=%
\mathcal{B}_{B}$ are extermal NUT solutions provided the mass parameter is
\begin{eqnarray}
m_{n}^{A} &=&-\frac{128\Lambda N^{7}}{105},  \label{mttt8} \\
m_{n}^{B} &=&-\frac{16N^{5}}{105}(8\Lambda N^{2}+7)  \label{mstt8}
\end{eqnarray}
These results for eight-dimensional Gauss-Bonnet gravity are consistent with
our conjectures. Again, one may note that the former extremal NUT solution
does not have a curvature singularity at $r=N$ whereas the latter does.

\subsection{Taub-Bolt Solutions}

The conditions for having a regular bolt solution are (i) $F(r=r_{b})=0$ and
$F^{\prime }(r_{b})=1/(4N)$ with $r_{b}>N$. Condition (ii) again follows
from the fact that we want to avoid a conical singularity at the bolt,
together with the fact that the period of $\tau $ will still be $16\pi N$.
Now applying these conditions for $\mathcal{B}=\mathbb{CP}^{3}$ gives the
following equation for $r_{b}$:
\begin{equation}
4N\Lambda {r_{b}}^{4}+3{r_{b}}^{3}-4(3N+2\Lambda N^{3}){r_{b}}^{2}+3(8\alpha
-N^{2})r_{b}+4N(\Lambda N^{4}+3N^{2}-9\alpha )=0  \label{bmatcp8}
\end{equation}
where again one has two bolt solutions for negative $\Lambda $ provided $%
N<N_{\text{max}}$ where $N_{\text{max}}$\ is the smaller root of the
following equation
\begin{eqnarray}
&&-320\Lambda ^{2}(27+64\Lambda \alpha )(15+64\Lambda \alpha
)N^{10}-2880\Lambda (9+16\Lambda \alpha )(15+64\Lambda \alpha
)N^{8}  \notag
\\
&&+9(1179648\Lambda ^{3}\alpha ^{3}+110182\Lambda ^{2}\alpha
^{2}+185760\Lambda \alpha -2025)N^{6}  \notag \\
&&-216\alpha (43008\Lambda ^{2}\alpha ^{2}+8992\Lambda \alpha -375)N^{4}
\notag \\
&&+324\alpha ^{2}(12288\Lambda ^{2}\alpha ^{2}+2688\Lambda \alpha
-301)N^{2}+41472\alpha ^{3}=0  \label{R8cp3}
\end{eqnarray}
As $\Lambda $ goes to zero Eq. (\ref{bmatcp8}) becomes
\begin{equation}
{r_{b}}^{3}-4N{r_{b}}^{2}+(8\alpha -N^{2})r_{b}+4N(N^{2}-3\alpha )=0
\label{E8cp3}
\end{equation}
which holds for locally asymptotically flat solutions. For all N, there is
only one solution for which $r_b>N$.

Taub-bolt solutions for the case of $\mathcal{B}=S^{2}\times \mathbb{CP}^{2}$
exist provided $N\leq N_{\mathrm{\max }}$ where $N_{\mathrm{\max }}$ is now
given by the smallest real root of the following equation
\begin{eqnarray}
&&-4800\Lambda ^{2}(243+896\Lambda \alpha )(27+128\Lambda \alpha
)N^{10}-2880\Lambda (499712\Lambda ^{2}\alpha ^{2}+1260496\Lambda
\alpha
+32805)N^{8}  \notag \\
&&+(5532286976\Lambda ^{3}\alpha ^{3}+3243829248\Lambda ^{2}\alpha
^{2}+422703360\Lambda \alpha -4428675)N^{6}  \notag \\
&&-72\alpha (3381248\Lambda ^{2}\alpha ^{2}+623232\Lambda \alpha -32265)N^{4}
\notag \\
&&+139968\alpha ^{2}(6912\Lambda ^{2}\alpha ^{2}+1472\Lambda
\alpha -159)N^{2}+10077696\alpha ^{3}=0
\end{eqnarray}
In this case there exist two $r_{b}$'s which are the real roots of
\begin{equation}
4N\Lambda {r_{b}}^{4}+3{r_{b}}^{3}-4(3N+2\Lambda N^{3}){r_{b}}^{2}+3(8\alpha
-N^{2})r_{b}+4N(\Lambda N^{4}+3N^{2}-\frac{32}{3}\alpha )=0  \label{s2cp2reg}
\end{equation}

Next we consider the Taub-bolt solutions for $\mathcal{B}=S^{2}\times
S^{2}\times S^{2}$. One finds that $r_{b}$ is given by the solution of the
following equation

\begin{equation}
4N\Lambda {r_{b}}^{4}+3{r_{b}}^{3}-4(3N+2\Lambda N^{3}){r_{b}}^{2}+3(8\alpha
-N^{2})r_{b}+4N(\Lambda N^{4}+3N^{2}-12\alpha )=0
\end{equation}
One has two bolt solutions for $N<N_{\mathrm{\max }}$ and one solution for $%
N=N_{\mathrm{\max }}$ where $N_{\mathrm{\max }}$ is now given by the
smallest real root of the following equation
\begin{eqnarray}
&&-64\Lambda ^{2}(27+128\Lambda \alpha )(25+128\Lambda \alpha
)N^{10}-44\Lambda (45056\Lambda ^{2}\alpha ^{2}+19152\Lambda
\alpha
+2025)N^{8}  \notag \\
&&+(12058624\Lambda ^{3}\alpha ^{3}+5548032\Lambda ^{2}\alpha
^{2}+597888\Lambda \alpha -6075)N^{6}  \notag \\
&&-72\alpha (38912\Lambda ^{2}\alpha ^{2}+6144\Lambda \alpha -417)N^{4}
\notag \\
&&+27648\alpha ^{2}(48\Lambda ^{2}\alpha ^{2}+10\Lambda \alpha
-1)N^{2}+13824\alpha ^{3}=0
\end{eqnarray}

For the locally asymptotic flat case, $r_{b}$ is the solution to the
equation
\begin{equation}
{r_{b}}^{3}-4N{r_{b}}^{2}+(8\alpha -N^{2})r_{b}+12N(N^{2}-4\alpha )=0
\label{8dbolts2cubed}
\end{equation}
and using reasoning similar to the 6-dimensional case we have only one bolt
solution.

For the case of $\mathcal{B}=T^{2}\times T^{2}\times T^{2}$ and $\mathcal{B}%
=T^{2}\times T^{2}\times S^{2}$, Euclidean regularity at the bolt requires
the period of $\tau $ to be
\begin{equation}
\beta =-\frac{12\pi r_{b}}{\Lambda ({r_{b}}^{2}-N^{2})}
\end{equation}
and
\begin{equation}
\beta =\frac{4\pi r_{b}(3{r_{b}}^{2}-3N^{2}+8\alpha )}{({r_{b}}%
^{2}-N^{2})[1-\Lambda ({r_{b}}^{2}-N^{2})]}
\end{equation}
respectively. As $r_{b}$ varies from $N$ to infinity, one covers the whole
temperature range from $0$ to $\infty $, and therefore one can have bolt
solutions. Again, one may note that for the case of asymptotic locally flat
solution with base space $\mathcal{B}=T^{2}\times T^{2}\times T^{2}$, there
is no black hole solution.

As with the six-dimensional case, there is a maximum value for $\alpha $ for
all base spaces that ensures all bolt and NUT solutions are regular in eight
dimensions (as in the case of static solutions).

\section{Ten-dimensional Solutions\label{10d}}

In ten dimensions there are more possibilities for the base space $\mathcal{B%
}$. It can be an $8$-dimensional space, the product of a $6$-dimensional
space with a $2$-dimensional one, a product of two $4$-dimensional spaces, a
product of a $4$-dimensional space with two $2$-dimensional spaces, or the
product of four $2$-dimensional spaces. In all of these cases the form of
the function $F(r)$ is

\begin{eqnarray}
F(r) &=&\frac{(r^{2}-N^{2})^{2}}{12\alpha (7r^{2}+3N^{2})}\left( 1+\frac{%
3p\alpha }{2(r^{2}-N^{2})}-\sqrt{B(r)+C(r)}\right) ,  \notag \\
B(r) &=&1+\frac{36\alpha mr\left( 7r^{2}+3N^{2}\right) }{(r^{2}-N^{2})^{6}}+%
\frac{6p\alpha N^{2}}{35(r^{2}-N^{2})^{6}}%
(3r^{8}-28N^{2}r^{6}+210N^{4}r^{4}+420N^{6}r^{2}+35N^{8})  \notag \\
&&+\frac{2\alpha \Lambda \left( 7r^{2}+3N^{2}\right) }{21(r^{2}-N^{2})^{6}}%
(7r^{10}-45N^{2}r^{8}-126N^{4}r^{6}-210N^{6}r^{4}+315N^{8}r^{2}+63N^{10})
\label{F10}
\end{eqnarray}
where $p$ is the dimensionality of the curved portion of the base space, and
the function $C(r)$ depends on the choice of the base space $\mathcal{B}$.

The first case is when one chooses the $8$-dimensional base space to be $%
\mathbb{CP}^{4}$. In this case the one-form $A$ and the metric of $\mathbb{CP%
}^{4}$ may be written as:
\begin{eqnarray}
A_4&=&10\sin^2 \xi_4 \left(d\psi_4+ \frac{1}{8}A_3\right)
\label{A4} \\
d{\Sigma _{3}}^{2} &=&10\{d{\xi_4}^2+\sin^2 \xi_4 \cos^2 \xi_4 (d\psi_4+%
\frac{1}{8}A_3)^2 +\frac{1}{8}\sin^2\xi_4 d{\Sigma _{3}}^{2} \}
\label{CP4}
\end{eqnarray}
where $A_3$ and $d{\Sigma _{3}}^{2}$ are given by Eqs. (\ref{A3}) and (\ref
{CP3})

The 1-forms for all the other cases have been introduced in the
previous sections, so we give only the function $C(r)$ for
different base spaces in the following table:\newline

\begin{tabular}{|c|c|}
\hline
$\mathcal{B}$ & $(r^{2}-N^{2})^{6}C(r)/\alpha ^{2}$ \\ \hline
$\,{\mathbb{CP}^{4}}$ & $-\frac{144}{25}%
(3r^{8}-28N^{2}r^{6}+210N^{4}r^{4}+420N^{6}r^{2}+35N^{8})$ \\ \hline
$T^{2}\times T^{2}\times T^{2}\times T^{2}$ & $0$ \\ \hline
$T^{2}\times T^{2}\times T^{2}\times S^{2}$ & $9(r^{2}-N^{2})^{4}$ \\ \hline
$T^{2}\times T^{2}\times S^{2}\times S^{2}$ & $\frac{12}{5}%
(r^{8}+4N^{2}r^{6}-90N^{4}r^{4}-220N^{6}r^{2}-15N^{8})$ \\ \hline
$T^{2}\times S^{2}\times S^{2}\times S^{2}$ & $-\frac{9}{5}%
(11r^{8}-76N^{2}r^{6}+450N^{4}r^{4}+820N^{6}r^{2}+75N^{8})$ \\ \hline
$S^{2}\times S^{2}\times S^{2}\times S^{2}$ & -$\frac{285}{5}%
(r^{8}-6N^{2}r^{6}+300N^{4}r^{4}+50N^{6}r^{2}+5N^{8})$ \\ \hline
$T^{2}\times \mathbb{CP}^{3}$ & $\frac{27}{5}%
(r^{8}+4N^{2}r^{6}-90N^{4}r^{4}-220N^{6}r^{2}-15N^{8})$ \\ \hline
$S^{2}\times \mathbb{CP}^{3}$ & $-\frac{18}{5}%
(9r^{8}-64N^{2}r^{6}+390N^{4}r^{4}+720N^{6}r^{2}+65N^{8})$ \\ \hline
$T^{2}\times T^{2}\times \mathbb{CP}^{2}$ & $\frac{4}{5}%
(17r^{8}-52N^{2}r^{6}-90N^{4}r^{4}-500N^{6}r^{2}-15N^{8})$ \\ \hline
$T^{2}\times S^{2}\times \mathbb{CP}^{2}$ & $-\frac{1}{5}%
(43r^{8}-428N^{2}r^{6}+3330N^{4}r^{4}+6740N^{6}r^{2}+555N^{8})$ \\ \hline
$S^{2}\times S^{2}\times \mathbb{CP}^{2}$ & $-\frac{8}{5}%
(29r^{8}-184N^{2}r^{6}+990N^{4}r^{4}+1720N^{6}r^{2}+165N^{8})$ \\ \hline
$\mathbb{CP}^{2}\times \mathbb{CP}^{2}$ & $-\frac{16}{5}%
(11r^{8}-76N^{2}r^{6}+450N^{4}r^{4}+820N^{6}r^{2}+75N^{8})$ \\ \hline
\end{tabular}
\newline

\bigskip

Note that the asymptotic behavior of all of these solutions is locally AdS
for $\Lambda <0$ provided $\left| \Lambda \right| <9/(42\alpha )$, locally
dS for $\Lambda >0$ and locally flat for $\Lambda =0$. As with the 6 and 8
dimensional cases, all the different $F(r)$'s have the same form as $\alpha $
goes to zero and reduce to the solutions of Einstein gravity.

\subsection{Taub-NUT Solutions}

Using the conditions for NUT solutions, (i) $F(r=N)=0$ and (ii) $F^{\prime
}(r=N)=1/(5N)$, we find that Gauss-Bonnet gravity in ten dimensions admits
non-extreme NUT solutions only when the base space is chosen to be $\mathbb{%
CP}^{4}$. There is no curvature singularity at $r=N$ for this solution
provided the mass parameter is fixed to be
\begin{equation}
m_{n}=-\frac{128N^{5}}{4725}(25\Lambda N^{4}+90N^{2}-378\alpha )
\label{mcp4}
\end{equation}
On the other hand, the solutions with $\mathcal{B}=T^{2}\times T^{2}\times
T^{2}\times T^{2}=\mathcal{B}_{A}$ and $\mathcal{B}=T^{2}\times T^{2}\times
T^{2}\times S^{2}=\mathcal{B}_{B}$ are extermal NUT solution provided the
mass parameter is
\begin{eqnarray}
m_{n}^{A} &=&-\frac{128\Lambda N^{9}}{189},  \label{mttt10} \\
m_{n}^{B} &=&-\frac{64N^{7}}{945}(10\Lambda N^{2}+9)  \label{mstt10}
\end{eqnarray}
These results for ten-dimensional Gauss-Bonnet gravity are consistent with
our conjectures -- we have found that all other spaces with 2-dimensional
factors in the base yield singular solutions. It is also straightforward to
show that the former extremal NUT solution has no curvature singularity at $%
r=N$, whereas the latter has\textbf{.}

\subsection{Taub-Bolt Solutions}

The conditions for having a regular bolt solution are (i) $F(r=r_{b})=0$ and
$F^{\prime }(r_{b})=1/(5N)$ with $r_{b}>N$. Now applying these conditions
for $\mathcal{B}=\mathbb{CP}^{4}$ gives the following equation for $r_{b}$:
\begin{equation}
5N\Lambda {r_{b}}^{4}+4{r_{b}}^{3}-10N(2+2\Lambda N^{2}){r_{b}}%
^{2}+4(12\alpha -N^{2})r_{b}+N(5\Lambda N^{4}+20N^{2}-96\alpha )=0
\end{equation}
where there exists a maximum value for the NUT parameter given by the
smallest real root of the following equation
\begin{eqnarray}
&&-25\Lambda ^{2}(4+15\Lambda \alpha )(1+15\Lambda \alpha
)N^{10}-50\Lambda (8+27\Lambda \alpha )(1+15\Lambda \alpha )N^{8}  \notag \\
&&+(34500\Lambda ^{3}\alpha ^{3}+18875\Lambda ^{2}\alpha ^{2}+2320\Lambda
\alpha -16)N^{6}  \notag \\
&&-2\alpha (6975\Lambda ^{2}\alpha ^{2}+1205\Lambda \alpha -48)N^{4}  \notag
\\
&&+\alpha ^{2}(6075\Lambda ^{2}\alpha ^{2}+1170\Lambda \alpha
-109)N^{2}+48\alpha ^{3}=0
\end{eqnarray}
in order for bolt solutions to exist. For the locally asymptotic flat case, $%
r_{b}$ is the solution of
\begin{equation}
{r_{b}}^{3}-5N{r_{b}}^{2}+(12\alpha -N^{2})r_{b}+N(5N^{2}-24\alpha )=0
\label{10dbolt}
\end{equation}
which yields only a single bolt solution.

For the case of $\mathcal{B}=T^{2}\times T^{2}\times T^{2}\times T^{2}$ and $%
\mathcal{B}=S^{2}\times T^{2}\times T^{2}\times T^{2}$, Euclidean regularity
at the bolt requires the period of $\tau $ to be
\begin{equation}
\beta =-\frac{16\pi r_{b}}{\Lambda ({r_{b}}^{2}-N^{2})}
\end{equation}
and
\begin{equation}
\beta =\frac{16\pi r_{b}({r_{b}}^{2}-N^{2}+3\alpha )}{({r_{b}}%
^{2}-N^{2})[1-\Lambda ({r_{b}}^{2}-N^{2})]}
\end{equation}
respectively. As $r_{b}$ varies from $N$ to infinity, one covers the whole
temperature range from $0$ to $\infty $, and therefore one can have bolt
solutions. Again, one may note that for the case of asymptotic locally flat
solution with base space $\mathcal{B}=T^{2}\times T^{2}\times T^{2}$, there
is no black hole solution.

Gauss-Bonnet gravity has other Taub-bolt solutions in ten dimensions
corresponding to the various base space factors given in the above table. We
shall omit the details of these solutions here, and simply note that it is
sufficient to apply the regularity and reality conditions for bolt solutions
mentioned at the beginning of this subsection to find the equation for%
\textbf{\ }$r_{b}$.

\section{The General Form of the Solutions in $d$ Dimensions\label{dd}}

In this section we write down the solutions of field equation (\ref{Geq})
for the metric (\ref{TN}). We find that the form of $F(r)$ in any dimension
and for any choice of the base space is
\begin{equation}
F(r)=\frac{(r^{2}-N^{2})\{r^{2}-N^{2}+\frac{2(d-4)}{d-2}p\alpha \}}{%
2(d-4)\alpha \{(d-3)r^{2}+3N^{2}\}}-\sqrt{F_{2}(r)}  \label{Fd}
\end{equation}
where $p$ is the sum of the dimensions of the curved factor spaces of $%
\mathcal{B}$.

Here, we consider only the NUT/bolt solutions of those cases for
which Gauss-Bonnet gravity admits NUT solutions, leaving out the
other cases which have only bolt solution for reasons of
economy\textbf{.} Our conjecture implies that there are three
cases that have NUT solutions in $2k+2$ dimensions.

\subsection{Taub-NUT/bolt solutions for the base space $\mathcal{B}=\mathbb{%
CP}^{k}$}

The only case for which Gauss-Bonnet gravity admits non-extreme NUT solutions in $%
2k+2$ dimensions is when the base space is $\mathcal{B}=\mathbb{CP}^{k}$. In
this case the metric may be written as \cite{Pop}
\begin{equation}
d\Sigma _{k}^{2}=(2k+2)\left\{ d\xi _{k}^{2}+\sin ^{2}\xi _{k}\cos
^{2}\xi _{k}(d\psi _{k}+\frac{1}{2k}A_{k-1})^{2}+\frac{1}{2k}\sin
^{2}\xi _{k}d\Sigma _{k-1}^{2}\right\}  \label{CPk}
\end{equation}
where $A_{k-1}$\ is the K\"{a}hler potential of $\mathbb{CP}^{k-1}$. Here $%
\xi _{k}$\ and $\psi _{k}$\ are the extra coordinates corresponding to $%
\mathbb{CP}^{k}$ with respect to $\mathbb{CP}^{k-1}$. Also, the metric is
normalized such that, Ricci tensor is equal to the metric, $R_{\mu \nu
}=g_{\mu \nu }$. The 1-form $A_{k}$, which is the K\"{a}hler potential of $%
\mathbb{CP}^{k}$, is
\begin{equation}
A_{k}=(2k+2)\sin ^{2}\xi _{k}(d\psi _{k}+\frac{1}{2k}A_{k-1})
\label{Ak}
\end{equation}
Now the $tt$-component of the field equation can be written as:
\begin{equation}
\Gamma _{1}rF^{\prime }(r)+\Gamma _{2}F^{2}(r)+\Gamma _{3}F(r)+\Gamma _{4}=0,
\label{ECP}
\end{equation}
where $\Gamma _{1}$, $\Gamma _{2}$, $\Gamma _{3}$ and $\Gamma _{4}$ are
\begin{eqnarray}
\Gamma _{1} &=&-\alpha \left( (d-3)r^{2}+3N^{2}\right)
F(r)+(r^{2}-N^{2})\left( \alpha +\frac{r^{2}-N^{2}}{2(d-4)}\right) ,  \notag
\\
\Gamma _{2} &=&-\frac{\alpha }{2(r^{2}-N^{2})}\left\{
(d-3)(d-5)r^{4}+2(d-9)N^{2}r^{2}+3N^{4}\right\} ,  \notag \\
\Gamma _{3} &=&-\frac{(r^{2}-N^{2})[(d-3)r^{2}+N^{2}]}{2(d-4)}+\alpha \left(
(d-5)r^{2}+N^{2}\right) ,  \notag \\
\Gamma _{4} &=&(r^{2}-N^{2})\left\{ \Lambda \frac{(r^{2}-N^{2})^{2}}{%
(d-2)(d-4)}-\frac{(r^{2}-N^{2})}{2(d-4)}-\frac{(d-2)\alpha }{2d}\right\}
\label{GamCP}
\end{eqnarray}

We can write the general form of $F_{2}(r)$ in $d$ dimensions. Inserting the
general form of $F(r)$ given in Eq. (\ref{Fd}) into the differential
equation (\ref{ECP}), we find $F_{2}(r)$ for $\mathcal{B}=\mathbb{CP}^{k}$\
as:

\begin{equation}
F_{2}(r)=\frac{r\left( (-1)^{(d-2)/2}m+\frac{1}{d(d-2)(d-4)^{2}\alpha }\int
\Upsilon (r)dr\right) }{4\alpha \{(d-3)r^{2}+3N^{2}\}(r^{2}-N^{2})^{(d-6)/2}}
\label{F2cp}
\end{equation}
where $\Upsilon (r)$ is
\begin{eqnarray}
\Upsilon (r) &=&\frac{(r^{2}-N^{2})^{d/2}}{r^{2}[(d-3)r^{2}+3N^{2}]^{2}}%
\{-24(d-2)(d-3)(d-4)^{2}\alpha ^{2}+8d(d-2)\alpha \Lambda \lbrack
(d-3)r^{2}+3N^{2}]^{2}  \notag \\
&&+24d(d-2)(d-4)\alpha
N^{2}+d(d-2)[(d-1)(d-3)r^{4}+6(d-1)r^{2}N^{2}+3N^{4}]\}
\end{eqnarray}

The solutions given by Eqs. (\ref{Fd}) and (\ref{F2cp}) yield a NUT solution
for any given (even) dimension $d>4$ provided the mass parameter $m$ is
fixed to be
\begin{equation}
m_{\mathrm{nut}}=-\frac{(\frac{d}{2}-3)!2^{d/2}N^{d-5}}{2d(d-1)!!}\left\{
2d^{2}\Lambda N^{4}+d(d-1)(d-2)N^{2}-(d-1)(d-2)(d-3)(d-4)\alpha
\right\} \label{mncp}
\end{equation}
This solution has no curvature singularity at $r=N$.

Solutions given by Eqs. (\ref{Fd}) and (\ref{F2cp}) for $m\neq
m_{\mathrm{nut }}$ in any dimension can be regarded as bolt
solutions. The value of the bolt
radius $r_{b}>N$ may be found from the regularity conditions (i) $%
F(r=r_{b})=0$ and $F^{\prime }(r_{b})=2/(dN)$. Applying these for $\mathcal{B%
}=\mathbb{CP}^{k}$ gives the following equation for $r_{b}$:
\begin{eqnarray}
0 &=&2d\Lambda N{r_{b}}^{4}+2(d-2){r_{b}}^{3}-dN\left[ (d-2)+4\Lambda N^{2}%
\right] {r_{b}}^{2}+2(d-2)\left[ 2(d-4)\alpha -N^{2}\right] r_{b}  \notag \\
&&+N\left[ 2d\Lambda N^{4}+d(d-2)N^{2}-(d-4)(d-2)^{2}\alpha \right]
\end{eqnarray}
which has two real roots larger than $N$, provided $N<N_{\mathrm{\max }}$
and one for $N=N_{\mathrm{\max }}$, where $N_{\mathrm{\max }}$ is the
smaller real root of the following equation
\begin{eqnarray}
&&-4(d+2)d^{2}[8d(d-4)\alpha \Lambda +d^{2}-4][8d(d-4)(d-6)\alpha
\Lambda
+d^{3}-6d^{2}+12d-8]\Lambda ^{2}N^{10}  \notag \\
&&-2d^{2}(d-2)(d+2)[8d(d-4)\alpha \Lambda +d^{2}-4][8(d-4)(d^{2}-6d-4)\alpha
\Lambda +d^{3}-6d^{2}+12d-8]\Lambda N^{8}  \notag \\
&&+(d-2)^{2}\{128d^{3}(d-4)^{3}(d^{2}-4d-14)\alpha ^{3}\Lambda
^{3}+32d^{2}(d-4)^{2}(d^{4}-4d^{3}-4d^{2}+40d+40)\Lambda ^{2}\alpha ^{2}
\notag \\
&&+2d(d-4)(d^{2}-4)(d^{4}-4d^{3}+6d^{2}+80d+24)\alpha \Lambda
-d^{6}+4d^{5}+4d^{4}-32d^{3}+16d^{2}+64d-64\}N^{6}  \notag \\
&&-(d-4)(d-2)^{2}\{192d^{2}(d-4)^{2}(d^{2}+2d+4)\alpha ^{2}\Lambda
^{2}+16d(d-2)(d-4)(d^{3}+8d^{2}+14d-12)\alpha \Lambda  \notag \\
&&-d^{6}+6d^{5}-16d^{4}-48d^{3}+176d^{2}+96d-384\}\alpha
N^{4}+(d-4)^{2}(d-2)^{2}\{1728d^{2}(d-4)^{2}\alpha ^{2}\Lambda ^{2}  \notag
\\
&&+48d(d-2)(5d^{2}-18d-8)\alpha \Lambda
-13d^{4}+16d^{3}+8d^{2}+192d-336\}\alpha ^{2}N^{2}\notag
\\
&& +128(d-2)^{4}(d-4)^{3}\alpha ^{3}=0
\end{eqnarray}

\bigskip

For the locally asymptotically flat case, $r_{b}$ is the solution
of
\begin{equation}
2{r_{b}}^{3}-dN{r_{b}}^{2}+2\left[ 2(d-4)\alpha -N^{2}\right] r_{b}+N\left[
dN^{2}-(d-4)(d-2)\alpha \right] =0  \label{dbolt}
\end{equation}
where there is no condition in order for a bolt solution to exist,
except in six dimensions for which there exists a lower limit of
the NUT charge.

Next we write down the general form of the solutions with the base space $%
\mathcal{B}=T^{2}\times ...\times T^{2}$. The field equation is given by (%
\ref{ECP}), where now
\begin{eqnarray}
\Gamma _{1} &=&-\alpha \{(d-3)r^{2}+3N^{2}\}F+\frac{(r^{2}-N^{2})^{2}}{2(d-4)%
},  \notag \\
\Gamma _{2} &=&-\frac{\alpha }{2(r^{2}-N^{2})}\left\{
(d-3)(d-5)r^{4}+2(d-9)N^{2}r^{2}+3N^{4}\right\} ,  \notag \\
\Gamma _{3} &=&-\frac{(r^{2}-N^{2})[(d-3)r^{2}+N^{2}]}{2(d-4)},  \notag \\
\Gamma _{4} &=&\Lambda \frac{(r^{2}-N^{2})^{3}}{(d-2)(d-4)}  \label{GamTT}
\end{eqnarray}
Inserting the general form of $F(r)$ given in Eq. (\ref{Fd}) into the
differential equation (\ref{ECP}) with $\Gamma _{i}$'s of Eqs. (\ref{GamTT}%
), we find that $F_{2}(r)$ for $\mathcal{B}=T^{2}\times T^{2}\times
...\times T^{2}$ is
\begin{equation}
F_{2}(r)=\frac{r\left( (-1)^{(d-2)/2}m+\frac{1}{(d-2)(d-4)^{2}\alpha }\int
\Upsilon (r)dr\right) }{4\alpha \{(d-3)r^{2}+3N^{2}\}(r^{2}-N^{2})^{(d-6)/2}}
\label{F2TT}
\end{equation}
where $\Upsilon (r)$ is
\begin{eqnarray*}
\Upsilon (r) &=&\frac{(r^{2}-N^{2})^{d/2}}{r^{2}\{(d-3)r^{2}+3N^{2}\}^{2}}%
\{8(d-4)\alpha \Lambda \lbrack (d-3)r^{2}+3N^{2}]^{2} \\
&&+(d-2)[(d-1)(d-3)r^{4}+6(d-1)r^{2}N^{2}+3N^{4}]
\end{eqnarray*}

The solutions given by Eqs. (\ref{Fd}) and (\ref{F2TT}) yield a NUT solution
for any given (even) dimension $d>4$ provided the mass parameter $m$ is
fixed to be
\begin{equation}
m_{\mathrm{nut }}=-\frac{d(\frac{d}{2}-3)!2^{d/2}}{(d-1)!!}\Lambda
N^{d-1} \label{mnTT}
\end{equation}
where in this case the spacetime has no curvature singularity at
$r=N$. Also one may find that Euclidean regularity at the bolt
requires the period of $\tau $ to be
\begin{equation}
\beta =-\frac{2(d-2)\pi r_{b}}{\Lambda ({r_{b}}^{2}-N^{2})}  \label{betTT}
\end{equation}
and can have any value from zero to infinity as $r_{b}$ varies from $N$ to
infinity, and therefore one can have bolt solution.

Finally, we consider the solution when $\mathcal{B}=S^{2}\times T^{2}\times
...\times T^{2}$. In this case the field has the same form as Eq. (\ref{ECP}%
) with
\begin{eqnarray}
\Gamma _{1} &=&-\alpha \{(d-3)r^{2}+3N^{2}\}F+(r^{2}-N^{2})\{\frac{2}{d-2}%
\alpha +\frac{r^{2}-N^{2}}{2(d-4)}\},  \notag \\
\Gamma _{2} &=&-\frac{\alpha }{2(r^{2}-N^{2})}\left\{
(d-3)(d-5)r^{4}+2(d-9)N^{2}r^{2}+3N^{4}\right\} ,  \notag \\
\Gamma _{3} &=&-\frac{(r^{2}-N^{2})[(d-3)r^{2}+N^{2}]}{2(d-4)}+\frac{4(d-4)}{%
d-2}\alpha \{(d-5)r^{2}+N^{2}\},  \notag \\
\Gamma _{4} &=&\frac{(r^{2}-N^{2})^{2}[\Lambda (r^{2}-N^{2})-1]}{(d-2)(d-4)}
\label{GamST}
\end{eqnarray}

Inserting the general form of $F(r)$ given in Eq. (\ref{Fd}) into the
differential equation (\ref{ECP}) with $\Gamma _{i}$'s of Eqs. (\ref{GamST}%
), we find that $F_{2}(r)$ for $\mathcal{B}=T^{2}\times T^{2}\times
...\times T^{2}$ is
\begin{equation}
F_{2}(r)=\frac{r\left( (-1)^{(d-2)/2}m+\frac{1}{(d-2)^{2}(d-4)^{2}\alpha }%
\int \Upsilon (r)dr\right) }{4\alpha
\{(d-3)r^{2}+3N^{2}\}(r^{2}-N^{2})^{(d-6)/2}}  \label{F2ST}
\end{equation}
where $\Upsilon (r)$ is
\begin{eqnarray}
\Upsilon (r) &=&\frac{(r^{2}-N^{2})^{d/2}}{r^{2}\{(d-3)r^{2}+3N^{2}\}^{2}}%
\{16(d-4)^{2}\alpha ^{2}\frac{(d-3)(d-5)r^{4}+6(d-3)N^{2}r^{2}+3N^{4}}{%
(r^{2}-N^{2})^{2}}  \notag \\
&&+8(d-2)(d-4)\alpha \Lambda \lbrack
(d-3)r^{2}+3N^{2}]^{2}+48(d-2)(d-4)\alpha N^{2}  \notag \\
&&-(d-2)^{2}[(d-1)(d-3)r^{4}+6(d-1)r^{2}N^{2}+3N^{4}]\}
\end{eqnarray}

The solutions given by Eqs. (\ref{Fd}) and (\ref{F2ST}) yield a NUT solution
for any given (even) dimension $d\geq 4$ with curvature singularity at $r=N$%
, provided the mass parameter $m$ is fixed to be
\begin{equation}
m_{\mathrm{nut}}=-\frac{(\frac{d}{2}-3)!2^{d/2}}{(d-1)!!}N^{d-3}\left\{
d\Lambda N^{2}+(d-1)\right\}   \label{mnST}
\end{equation}

Also one may find that the Euclidean regularity at the bolt requires the
period of $\tau $ to be
\begin{equation}
\beta =\frac{2(d-2)\pi r_{b}({r_{b}}^{2}-N^{2}+\frac{4(d-4)}{d-2}\alpha )}{({%
r_{b}}^{2}-N^{2})[1-\Lambda ({r_{b}}^{2}-N^{2})]}  \label{betST}
\end{equation}
Again, $\beta $ of Eq. (\ref{betST}) can have any value from zero to
infinity as $r_{b}$ varies from $N$ to infinity, and therefore one can have
bolt solution.

The asymptotic behavior of all of these solutions is locally AdS for\ $%
\Lambda <0$ provided $\left| \Lambda \right|
<(d-1)(d-2)/[(d-3)(d-4)\alpha ]$ locally dS for $\Lambda >0$ and
locally flat for $\Lambda =0$ . All the different $F(r)$'s for
differing base spaces have the same form as $\alpha$ goes to zero
and reduce to the solutions of Einstein gravity.

\section{Concluding Remarks \label{con}}

We have considered the existence of Taub-NUT/bolt solutions in Gauss-Bonnet
gravity with and without cosmological term. These solutions are constructed
as circle fibrations over even dimensional spaces that in general are
products of Einstein-K\"{a}hler spaces. We found that the function $F(r)$ of
the metric depends on the specific form of the base factors on which one
constructs the circle fibration. In other words we found that the solutions
are sensitive to the geometry of the base space, in contrast to Einstein
gravity where the metric in any dimension is independent of the specific
form of the base factors. We restricted ourselves to the cases of base
spaces with zero or positive curvature factor spaces, since when the base
space has $2$-hyperboloids with negative curvature, then the function $F(r)$
is not real for the whole range of $r$ when $\alpha $ is positive.

We found that when Einstein gravity admits non-extremal NUT
solutions with no curvature singularity at $r=N$, then there
exists a non-extremal NUT solution in Gauss-Bonnet gravity. In
$(2k+2)$-dimensional
spacetime, this happens when the metric of the base space is chosen to be $%
\mathbb{CP}^{k}$. Indeed, Gauss-Bonnet gravity does not admit non-extreme
NUT solutions with any other base space. We also found that when the base
space has at most a 2-dimensional curved factor space with positive
curvature, then Gauss-Bonnet gravity admits extremal NUT solutions where the
temperature of the horizon at $r=N$ vanishes. We have extended these
observations to two conjectures about the existence of NUT solutions in
Gauss-Bonnet gravity.

We also found the bolt solutions of Gauss-Bonnet gravity in
various dimensions and different base spaces, and gave the
equations which can be solved for the horizon radius of the bolt
solution. For $\Lambda \neq 0$, there is a maximal value for the
NUT charge in terms of the parameters of the metric that $N$ must
be less than or equal to that in order to have bolt solutions. In
this case for $\Lambda <0$ we have two bolt solutions in
Gauss-Bonnet gravity if $N<N_{\mathrm{\max }}$ and one bolt
solution if $N=N_{\mathrm{\max }}$. The value of
$N_{\mathrm{max}}$ is larger in Gauss-Bonnet gravity than in
Einstein gravity. For the case of $\Lambda =0$ we always have only
one bolt solution. In the 6-dimensional case with the base space
$\mathbb{CP}^{2}$ the bolt solution exists provided the NUT charge
is larger than its minimal value of $\sqrt{\alpha }$. This
situation is quite unlike that in Einstein gravity, where the NUT
charge can smoothly approach zero. We note that Gauss-Bonnet
gravity in six dimensions gives the most general second order
differential equation in classical gravity, while in higher
dimensions in order to have the most general second order
differential equation, one should turn on other higher terms of
Lovelock gravity.

Our solutions have been generalized in an obvious way for even dimensions
higher than ten. We gave the differential equation and the general form of
the function $F(r)$ in any arbitrary even dimensions. For instance, in
twelve dimensions, the base space is ten-dimensional and in general can be
factorized as a $10$-dimensional space, the product of an $8$-dimensional
space with a $2$-dimensional one, a product of $6$-dimensional space with a $%
4$-dimensional space, a product of a $6$-dimensional space with two $2$%
-dimensional spaces, a product of two $4$-dimensional spaces with a $2$%
-dimensional space, a product of a $4$-dimensional spaces with three $2$%
-dimensional spaces, or the product of five $2$-dimensional spaces. In any
dimension $2k+2$, we have only one non-extremal NUT solution with $\mathbb{CP%
}^{k}$ as the base space, and two extremal NUT solutions with the
base spaces $T^{2}\times T^{2}\times .....\times T^{2}$ and
$S^{2}\times T^{2}\times T^{2}\times .....\times T^{2}$. There is
no curvature singularity for the first two case, while for the
latter case, the spacetime has curvature singularity at $r=N$.

Insofar as Gauss-Bonnet gravity is expected to model leading-order quantum
corrections to Einstein gravity, we see that quantum effects can be expected
to single out a preferred non-singular base space and to yield a minimal
value for the NUT charge in six dimensions. Thermodynamically this
corresponds to a maximal temperature. The study of thermodynamic properties
of these solutions, the investigation of the existence of NUT solutions in
continued Lovelock gravity, or Lovelock gravity with higher order terms
remain to be carried out in future.\newline

{\Large \textbf{Acknowledgements}}

We are grateful to C. Stelea for helpful discussions about $\mathbb{CP}^{k}$%
. M. H. D. would like to thank the University of Waterloo and the Perimeter
Institute for Theoretical Physics for support and hospitality during the
course of this work. This work was supported in part by the Natural Sciences
and Engineering Council of Canada.


\begin{thebibliography}{99}
\bibitem{Chamblin}  A.~Chamblin, R.~Emparan, C.~V.~Johnson and
R.~C.~Myers,``Large N phases, gravitational instantons and the NUTs and
bolts of AdS holography'', Phys. Rev. D \textbf{59}, 064010 (1999)
[arXiv:hep-th/9808177].

\bibitem{Hawking}  S.~W.~Hawking, C.~J.~Hunter and D.~N.~Page, ``Nut Charge,
Anti-de Sitter Space and Entropy'', Phys. Rev. D \textbf{59}, 044033 (1999)
[hep-th/9809035].

\bibitem{MannMisner}  R.~B.~Mann, ``Misner String Entropy", Phys. Rev. D
\textbf{60}, 104047 (1999) [hep-th/9903229].

\bibitem{mbrane}  S. Cherkis and A. Hashimoto, ``Supergravity Solution of
Intersecting Branes and AdS/CFT with Flavour", JHEP \textbf{0211}, 036
(2002) [hep-th/0210105]; R. Clarkson, A. M. Ghezelbash and R. B. Mann, ``New
Reducible M-brane Solutions in D=11 Supergravity", JHEP \textbf{0404}, 063
(2004) [hep-th/0404071]; R. Clarkson, A. M. Ghezelbash and R. B. Mann, ``New
Reducible 5-brane Solutions in M-Theory", JHEP \textbf{0408}, {025} (2004)
[hep-th/0405148].

\bibitem{Taub}  A.~H.~Taub, ``Empty Space-Times Admitting a Three Parameter
Group of Motions'', Annal. Math. \textbf{53}, 472 (1951).

\bibitem{NUT}  E.~Newman, L.~Tamburino, and T.~Unti, ``Empty-space
generalization of the Schwarzschild metric'', J. Math. Phys.
\textbf{4}, 915 (1963).

\bibitem{Bais}  F.~A.~Bais and P.~Batenburg, ``A New Class Of Higher
Dimensional Kaluza-Klein Monopole And Instanton Solutions'', Nucl. Phys. B
\textbf{253}, 162 (1985).

\bibitem{Page}  D.~N.~Page and C.~N.~Pope, ``Inhomogeneous Einstein Metrics
On Complex Line Bundles'', Class. Quant. Grav. \textbf{4}, 213 (1987).

\bibitem{Akbar}  M.~M.~Akbar and G.~W.~Gibbons, ``Ricci-flat metrics with
U(1) action and the Dirichlet boundary-value problem in Riemannian quantum
gravity and isoperimetric inequalities'', Class. Quant. Grav. \textbf{20},
1787 (2003) [arXiv:hep-th/0301026].

\bibitem{Robinson}  M.~M.~Taylor-Robinson, ``Higher dimensional Taub-Bolt
solutions and the entropy of non compact manifolds'' [hep-th/9809041].

\bibitem{Awad}  A.~Awad and A.~Chamblin, ``A bestiary of higher dimensional
Taub-NUT-AdS spacetimes'', Class. Quant. Grav. \textbf{19}, 2051 (2002)
[hep-th/0012240].

\bibitem{Lorenzo}  R.~Clarkson, L.~Fatibene and R.~B.~Mann, ``Thermodynamics
of ($d+1$)-dimensional NUT-charged AdS Spacetimes'', Nucl. Phys. \textbf{B652%
} 348 (2003) [hep-th/0210280].

\bibitem{Mann1}  Robert Mann and Cristian Stelea, ``Nuttier (a)dS black
holes in higher dimensions", Class. Quant. Grav. \textbf{21}, 2937 (2004).

\bibitem{Mann2}  Robert Mann and Cristian Stelea, New multiply Nutty
spacetimes" [hep-th/0508203].

\bibitem{Wit1}  M. B. Greens, J. H. Schwarz and E. Witten, \emph{Superstring
Theory}, (Cambridge University Press, Cambridge, England, 1987); D. Lust and
S. Theusen, \emph{Lectures on String Theory}, (Springer, Berlin, 1989); J.
Polchinski, \emph{String Theory}, (Cambridge University Press, Cambridge,
England, 1998).

\bibitem{Zw}  B. Zwiebach, ``Curvature squared terms and string theories",
Phys. Lett. \textbf{B156}, 315 (1985); B. Zumino, Phys. Rep.
``Gravity theories in more than four dimensions", \textbf{137},
109 (1986).

\bibitem{Lov}  D. Lovelock, ``The Einstein Tensor and Its Generalizations",
J. Math. Phys. \textbf{12}, 498 (1971); N. Deruelle and L. Farina-Busto,
``Lovelock gravitational field equations in cosmology", Phys. Rev. D \textbf{%
41}, 3696 (1990); G. A. Mena Marugan, ``Perturbative formalism of Lovelock
gravity", \emph{ibid}. \textbf{46}, 4320 (1992); 4340 (1992).

\bibitem{Des}  D. G. Boulware and S. Deser, Phys. Rev. Lett.,
``String-Generated Gravity Models", \textbf{55}, 2656 (1985); J. T. Wheeler,
``Symmetric solutions to the Gauss-Bonnet extended Einstein equations ",
Nucl. Phys. \textbf{B268}, 737 (1986).

\bibitem{Whe}  J. T. Wheeler, ``Symmetric solutions to the maximally
Gauss-Bonnet extended Einstein equations ", Nucl. Phys. \textbf{B273}, 732
(1986).

\bibitem{Deh1}  M. H. Dehghani and M. Shahmirzaie, ``Thermodynamics of
Asymptotic Flat Charged Black Holes in Third Order Lovelock
Gravity", Phys. Rev. D, \emph{accepted}, [hep-th/0506227].

\bibitem{Wil}  D. L. Wiltshire, ``Spherically symmetric solutions of
Einstein-Maxwell theory with a Gauss-Bonnet term", Phys. Lett. B \textbf{169}, 36 (1986).

\bibitem{Cai}  R. G. Cai, ``Gauss-Bonnet black holes in AdS spaces", Phys. Rev. D
\textbf{65}, 084014 (2002) [hep-th/0109133]; M. Cvetic, S. Nojiri,
S.D. Odintsov, ``Black hole thermodynamics and negative entropy in
de Sitter and anti-de Sitter Einstein Gauss-Bonnet gravity", Nucl.
Phys. \textbf{B628}, 295 (2002) [hep-th/0112045].

\bibitem{Deh2}  M. H. Dehghani, ``Charged rotating black branes in anti--de
Sitter Einstein-Gauss-Bonnet gravity"' Phys. Rev. D \textbf{67}, 064017
(2003) [hep-th/0211191]; ``Magnetic branes in Gauss-Bonnet gravity ", \emph{%
ibid}. \textbf{69}, 064024 (2004) [hep-th/0312030];
``Asymptotically (anti)-de Sitter solutions in Gauss-Bonnet
gravity without a cosmological constant, \emph{ibid}. \textbf{70},
064019 (2004) [hep-th/0405206].

\bibitem{Pop}  P. Hoxha, R. R. Martinez-Acosta, C. N. Pope, ``Kaluza-Klein
Consistency, Killing Vectors, and Kahler Spaces", Class. Quant. Grav.
\textbf{17}, 4207 (2000) [hep-th/0005172].
\end{thebibliography}
\end{document}